\documentclass[conference]{IEEEtran}
\IEEEoverridecommandlockouts

\usepackage{cite}
\usepackage{amsmath,amssymb,amsfonts}
\usepackage{algorithmic}
\usepackage{graphicx}
\usepackage{textcomp}
\usepackage{xcolor}
\usepackage{array}
\usepackage{booktabs}
\usepackage{mathtools}
\usepackage{listings}
\usepackage{tabularx}
\usepackage{url}
\usepackage{hyperref}
\usepackage{multirow}

\def\BibTeX{{\rm B\kern-.05em{\sc i\kern-.025em b}\kern-.08em
    T\kern-.1667em\lower.7ex\hbox{E}\kern-.125emX}}
\begin{document}

\title{Efficiency Hallucination: Formalizing and Measuring Behavioral Calibration in LLM-Based Code Optimization}

\author{\IEEEauthorblockN{Sarah Wilson{*}\thanks{{*}Current affiliation: Google.}}
\IEEEauthorblockA{
\textit{Columbia University}\\
sw4104@columbia.edu}
\and
\IEEEauthorblockN{Gail Kaiser}
\IEEEauthorblockA{
\textit{Columbia University}\\
kaiser@cs.columbia.edu}
\and
\IEEEauthorblockN{Patrick Musau}
\IEEEauthorblockA{
\textit{Vanderbilt University}\\
patrickmusau@vanderbilt.edu}

}

\maketitle

\begin{abstract}

The integration of Large Language Models (LLMs) into automated code optimization introduces a critical reliability risk we term the \textit{Efficiency Hallucination}: an LLM’s tendency to issue non-functional mutations with unsubstantiated performance claims on already-optimized code. This is driven by the \textit{Evaluation Trap}, wherein binary benchmarks incentivize unnecessary modifications over safely abstaining. We present a validation framework using classification penalty methods, evaluated across 180 optimization runs on nine models (GPT, Claude, Gemini) using EffiBench. Under standard prompts, models exhibit a 100\% over-edit rate on optimal code. Our guardrail raises correct abstention from 0\% to to 44.4\%, preserving a 100\% edit rate on sub-optimal code with zero false abstentions. Calibration is uneven: GPT-5.4 Mini approaches near-perfect abstention, and simple code is recognized more reliably than complex code. Our framework offers a training-free mechanism to mitigate LLM overconfidence before deployment in production.
\end{abstract}


\section{Introduction}
The automation of code optimization represents one of the highest-stakes applications of Large Language Models (LLMs) in software engineering. Classical compilers---from production toolchains like GCC~\cite{gcc} and LLVM/Clang~\cite{lattner2004llvm} to research systems such as CETUS~\cite{bae2013cetus} and PLUTO~\cite{uday2008apraticalautomaticpolynomialparallelizer}---operate through deterministic, rule-based static transformations. Modern LLM-based frameworks---AlphaEvolve~\cite{novikov2025alphaevolvecodingagentscientific}, LLaMoCo~\cite{ma2026llamaco}, PerfRL~\cite{duan2025perfrlsmalllanguagemodel}---go substantially further, leveraging deep semantic understanding to identify algorithmic improvements such as loop restructuring, data structure substitution, and closed-form mathematical replacements that static tools cannot discover. In some cases, these systems have surpassed prior human records in algorithmic problem solving~\cite{novikov2025alphaevolvecodingagentscientific}.

Yet this semantic fluency introduces a subtle and underexamined reliability failure. When presented with code that is already optimal, LLMs do not admit uncertainty: as our pilot study confirms (Section~\ref{sec:pilot}), they \textit{invent} optimizations. We term this the \textbf{Efficiency Hallucination}---a model's generation of plausible code edits accompanied by confident, unverifiable performance claims. Unlike functional hallucinations that cause test failures, Efficiency Hallucinations are largely invisible to per-commit test suites that gate everyday development---the code still compiles, passes tests, and may even contain comments confidently asserting a speedup, yet no actual performance improvement occurs. Even dedicated performance-regression suites, which typically run only before a release rather than per-commit, may not catch them until much later. In production pipelines, such edits cost review time and erode trust in automated tooling.

The root cause is what we call the \textbf{Evaluation Trap}. Optimization pipelines using benchmarks like PIE~\cite{shypula2024learning} and EffiBench~\cite{huang2025effibenchbenchmarkingefficiencyautomatically} evaluate success mainly via functional correctness and execution runtime, leaving unpenalized any hallucinated speedup claims a model generates in its textual rationale. This binary incentive creates an asymmetric risk structure: for a model encountering already-optimal code, admitting optimality yields zero optimization credit, while generating a plausible bluff is a zero-risk gamble that may capitalize on runtime variance. We formalize this mathematically through the Is-It-Valid (IIV) classification framework of Kalai et al.~\cite{kalai2025languagemodelshallucinate}, which demonstrates that generative error rates are mathematically guaranteed to be at least double the underlying misclassification rate.  Without an explicit penalty for over-optimization, bluffing becomes the dominant inference strategy for models encountering performance ceilings.

This paper makes four contributions. First, we formally define a taxonomy of Efficiency Hallucinations, distinguishing \textit{over-edits} (hallucinated optimizations on already-optimal code) from \textit{false abstentions} (failure to improve genuinely improvable code), and from the complementary true outcomes of correct edits and correct abstentions. Second, we provide the first application of the IIV framework to code optimization, explaining why current benchmark designs are mathematically biased toward elevated bluffing rates. Third, we introduce the \textbf{Optimal Baseline} evaluation methodology, which measures an agent's \textbf{Behavioral Calibration}---its wisdom to abstain---rather than raw optimization speedup. Fourth, we conduct a controlled 180-trial pilot study across nine models from three LLM families, validating these predictions empirically and revealing per-model, per-family, and per-problem variation with direct implications for the reliability engineering of AI-assisted software pipelines.

\section{Related Work}
\label{sec:related}
 
\subsection{LLM-Based Code Optimization}

The integration of LLMs into automated performance engineering spans three main directions: architectural specialization, agentic search strategies, and evaluation and benchmarking.

\textbf{Architectural Specialization.} The LLaMoCo framework~\cite{ma2026llamaco} addresses a core failure mode of general-purpose LLMs in optimization: their tendency toward API hallucinations, syntax errors, and hyper-parameter misinterpretations. LLaMoCo fine-tunes in two stages, first teaching the model to recognize what type of optimization problem it faces, then how to solve it---reducing the kind of confident misjudgment that, in our own results, surfaces as treating visually complex code as inherently improvable. The PerfRL framework~\cite{duan2025perfrlsmalllanguagemodel} fine-tunes a small model (CodeT5) with reinforcement learning from unit-test and compiler feedback; despite far fewer parameters, it matches or exceeds the larger CodeGen-2B baseline on speedup and runtime-reduction scores. These works demonstrate that specialization reduces the hallucination patterns above, motivating our investigation of whether it also improves behavioral calibration.

\textbf{Agentic Search Strategies.} 
AlphaEvolve~\cite{novikov2025alphaevolvecodingagentscientific} orchestrates autonomous evolutionary search over entire codebases, pairing high-throughput generation with strong-reasoner verification to discover improvements to established algorithms (e.g., a new bound for matrix multiplication). The CSE framework~\cite{hu2026controlledselfevolutionalgorithmiccode} adds hierarchical evolution memory for inter-task generalization, and Artemis~\cite{brookes2025evolvingexcellenceautomatedoptimization} automates agent configuration from execution logs. While these frameworks improve optimization capability, their iterative pressure creates the exact conditions under which Efficiency Hallucinations are most likely: models are incentivized to propose improvements in every round, even when the code has reached its ceiling.

\textbf{Evaluation and Benchmarking.} Recent work reveals a persistent gap between functional correctness and efficiency. Peng et al.~\cite{peng2025agentsperformcodeoptimization} find AI patches 18\% less likely than human patches to include performance validation, with some claiming extreme speedups (7400$\times$) without benchmarks; ENAMEL~\cite{qiu2025efficient} quantifies the gap (GPT-4: 0.831 functional pass rate, 0.454 efficiency score). SWE-Perf~\cite{he2025sweperflanguagemodelsoptimize} scales evaluation to 140 real GitHub performance PRs, yet---like ENAMEL---tests only genuinely improvable code; by construction it cannot surface the Evaluation Trap on already-optimal code, the failure mode we study here. Nogueira et al.~\cite{nogueirabeyond2025} confirm the same pattern in code generation---pass/fail metrics miss structural quality failures---making this cross-domain and motivating the Efficiency Hallucination.

\subsection{LLM Hallucination in Code}

CECoder~\cite{hececoder2025} reduces hallucinations in repository-level code generation through fine-grained code element retrieval; unlike retrieval-based grounding, our work targets the distinct failure mode where models understand the code but fabricate optimization improvements regardless. CodeMirage~\cite{agarwal2025codemiragehallucinationscodegenerated} identifies three code-specific hallucination subtypes: API misuse, logic errors, and efficiency misrepresentation (claiming superior complexity while producing sub-optimal code)---the last being most directly related to our work. Hallucinations in repository-level code are driven by cross-file dependency neglect, stale documentation retrieval, and over-generalization of local patterns~\cite{zhang2025llmhallucinationspracticalcode}. EffiBench~\cite{huang2025effibenchbenchmarkingefficiencyautomatically} specifically benchmarks efficiency claims, finding that even when LLMs produce correct solutions, they tend to use sub-optimal algorithms where human experts choose optimal ones.

The IIV framework of Kalai et al.~\cite{kalai2025languagemodelshallucinate} provides the mathematical foundation for this work. They demonstrate that text generation amplifies misclassification errors: if a model's misclassification rate on training facts is $\epsilon$, its generative error rate is at least $2\epsilon$---each misclassification compounds at generation because the model emits the incorrect pattern as if it were correct. At a 10\% misclassification rate, merging 1,000 LLM optimization PRs guarantees at least 200 unverified performance claims that may pass CI yet introduce latent complexity. This amplification is especially severe for patterns appearing only once in training data---singletons---because models cannot distinguish these from statistical noise. Optimization micro-patterns (e.g., a specific bitwise trick valid only for a narrow input range) are prototypical singletons, making Efficiency Hallucinations a mathematically predictable failure mode rather than a correctable bug.

\section{The Efficiency Hallucination Framework}
\label{sec:framework}

\subsection{The Evaluation Trap and IIV Formalization}

Following the IIV analysis~\cite{kalai2025languagemodelshallucinate}, we model the optimization decision as a binary classification problem over two example types: 

\begin{itemize}
  \item \textbf{Valid (+):} Edits that maintain functional correctness while achieving measurable
        runtime improvements.
  \item \textbf{Error (-):} Edits that maintain functional correctness but provide no runtime improvement, degrade performance, or introduce functional regressions.
\end{itemize}


To the best of our knowledge, current benchmarks provide no negative signal for the Error class during training or evaluation, removing the only mechanism that could suppress this mathematically guaranteed bluffing rate.

\subsection{Behavioral Calibration: Taxonomy}We define four disjoint outcome categories for any model response to an optimization request. Let $\mathcal{C}$ denote code at its performance ceiling (optimal) and $\mathcal{S}$ denote genuinely sub-optimal code.

\begin{itemize}
  \item \textbf{Edit (True Positive):} Model modifies $\mathcal{S}$. Correct outcome.
  \item \textbf{Abstention (True Negative):} Model signals $\mathcal{C}$ is already optimal
        (e.g., outputs \texttt{ALREADY\_OPTIMAL}). Correct outcome.
  \item \textbf{Over-Edit (False Positive):} Model modifies $\mathcal{C}$.
        This is an Efficiency Hallucination.
  \item \textbf{False Abstention (False Negative):} Model declines to improve $\mathcal{S}$.
        Pathological refusal.
\end{itemize}

A perfectly calibrated agent achieves edit rate $= 1.0$ on $\mathcal{S}$ and abstention rate $= 1.0$ on $\mathcal{C}$ simultaneously. We define the \textbf{Bluff Rate} as the over-edit frequency on optimal code, and the \textbf{False Abstention Rate} as the abstention frequency on sub-optimal code. Perfect calibration requires both to be zero. Under the Evaluation Trap, the Bluff Rate is guaranteed to be positive and approaches 1.0 as the performance ceiling becomes harder to verify statically.

\subsection{The Optimal Baseline Strategy}

Standard benchmarks measure capability by testing agents exclusively on improvable code. Our methodology adds a complementary evaluation axis: testing agents on \textit{top-percentile human solutions} from EffiBench~\cite{huang2025effibenchbenchmarkingefficiencyautomatically} that have effectively reached their theoretical performance ceiling. This Optimal Baseline design isolates the Bluff Rate from optimization capability, enabling independent measurement of each. We operationalize calibration via an IIV-derived \textbf{Penalty Prompt}:

\begin{quote}
  \textit{``Only suggest an edit if you are $>$90\% confident it improves execution speed; 
  otherwise, output \texttt{ALREADY\_OPTIMAL}.''}
\end{quote}

The threshold exploits a static-verification asymmetry: on sub-optimal code, a model can often confirm confidence exceeds 90\% from complexity analysis alone (replacing $O(n^2)$ with $O(n \log n)$ requires no runtime clock); on already-optimal code, no such static proof exists, so an honestly calibrated model must fall below threshold and abstain.

This explicit 90\% confidence threshold, grounded in the singleton-avoidance mechanism of the IIV framework, introduces a non-zero inference-time cost for the Error class---the missing signal that current benchmarks omit. By comparing model behavior under the control condition (no penalty, simply ``\textit{Optimize this code for execution speed.}'') versus the IIV penalty condition, we can directly measure whether models are capable of calibrated abstention and whether that capability degrades their optimization performance on improvable code.

\section{Pilot Study}
\label{sec:pilot}
 
\subsection{Experimental Setup}
We constructed a 180-trial dataset spanning two snippet types and two prompt conditions.

\textbf{Problem Pairs.} Five LeetCode problems were selected from the EffiBench benchmark suite~\cite{huang2025effibenchbenchmarkingefficiencyautomatically}, each yielding a paired optimal/sub-optimal snippet. The \textit{optimal} variant is the EffiBench top-percentile solution; the \textit{sub-optimal} variant is a functionally correct but algorithmically degraded version generated by Gemini-3.5-Flash and human-verified. The five pairs are: (1) Combination Sum II (optimal: sorted DFS with duplicate skipping; sub-optimal: brute-force DFS without pruning), (2) Remove Duplicates from Sorted Array II (optimal: $O(n)$ two-pointer; sub-optimal: $O(n^2)$ \texttt{list.count()} loop), (3) Is Same Tree (optimal: recursive structural comparison with short-circuit; sub-optimal: string serialization), (4) Finding 3-Digit Even Numbers (optimal: $O(1000)$ iteration with \texttt{Counter}; sub-optimal: $O(n^3)$ nested loops), and (5) Min Operations to Reduce an Integer to 0 (optimal: bit-manipulation; sub-optimal: naive step-by-step simulation). Using paired problems controls for problem-specific familiarity effects. All code and data are available at \url{https://github.com/sarah-wilsxn/efficiency-hallucinations}.

\textbf{Models and Conditions.} Nine models from three LLM families were evaluated via direct API---not through coding-agent wrappers (e.g., Claude Code, Codex CLI) whose refinement loops may partially suppress bluffing; agent evaluation is designated future work. Two conditions were used: \textit{control} (standard optimization request) and \textit{iiv\_penalty} (the Penalty Prompt from Section~\ref{sec:framework}). Model families: Claude (claude-opus-4.7-fast, claude-opus-4.8, claude-sonnet-4.5; $n=60$), Gemini (gemini-3-flash-preview, gemini-3.1-pro-preview, gemini-3.5-flash; $n=60$), and GPT (gpt-5-mini, gpt-5.4-mini, gpt-5.4; $n=60$). Each model received 20 trials: 5 optimal $\times$ 2 conditions + 5 sub-optimal $\times$ 2 conditions, yielding $9 \times 20 = 180$ total.

\subsection{Result I: The Evaluation Trap Confirmed}
Table~\ref{tab:condition_snippet} presents results disaggregated by prompt condition and snippet type. Under the control condition, the edit rate on optimal code is \textbf{100\%} across all four model families, with zero abstentions. Every model, in every trial, modified already-optimal code without hesitation, producing confident suggestions for changes that, by construction, should not improve execution speed. The edit rate on sub-optimal code is also 100\% under control, confirming that baseline optimization capability is uniformly high.

\begin{table}[t]
    \caption{Pilot Study Results by Prompt Condition and Snippet Type ($N = 180$)}
    \label{tab:condition_snippet}
    \centering
    \renewcommand{\arraystretch}{1.15}
    \begin{footnotesize}
    \begin{tabular}{@{}llcccc@{}}
    \toprule
    \textbf{Condition} & \textbf{Snippet} & \textbf{Edit} & \textbf{Abs.} &
    \textbf{O-Edit} & \textbf{F.Ab.} \\
    \midrule
    Control     & Optimal      & 1.000 & 0.000 & ---           & 0.000 \\
    Control     & Sub-Optimal  & 1.000 & 0.000 & ---           & 0.000 \\
    IIV Penalty & Optimal      & ---   & \textbf{0.444} & 0.556 & 0.000 \\
    IIV Penalty & Sub-Optimal  & 1.000 & 0.000 & 0.000         & 0.000 \\
    \bottomrule
    \end{tabular}
    \end{footnotesize}
    \smallskip

    {\small All values are proportions. Over-Edit and Edit are mutually exclusive by condition:
    any modification of optimal code under the penalty condition is classified as Over-Edit (not
    Edit). Over-Edit is undefined (---) under the control condition since no penalty is active.
    F.~Ab.\ $=$ False Abstention.}
\end{table}

This 100\% Bluff Rate confirms the Evaluation Trap: absent an explicit penalty, no tested model recognizes when code is already optimal, validating the IIV prediction that bluffing is the dominant strategy at the performance ceiling.

\subsection{Result II: Behavioral Calibration under IIV Penalty}

Introducing the Penalty Prompt produces a clear directional shift. On optimal code, the abstention rate rises from 0\% to \textbf{44.4\%} while the over-edit (Bluff) rate falls from 100\% to 55.6\%. Equally critical: the edit rate on sub-optimal code remains at \textbf{100\%} under penalty and the False Abstention Rate is \textbf{0.0\%} across all models. The IIV penalty does not cause models to refuse all edits: it specifically reduces over-editing on already-optimal code while leaving the 100\% edit rate on sub-optimal code intact. Calibration and capability are behaviorally independent.

The persistence of a 55.6\% Bluff Rate follows the IIV framework: the penalty asks a model to report confidence above 90\%, but that self-assessment \textit{is} the unreliable judgment IIV describes---the threshold changes the decision rule, not underlying calibration~\cite{kalai2025languagemodelshallucinate}. External runtime verification is the definitive solution, motivating our execution-based pipeline (Section~\ref{sec:conclusion}).

\subsection{Result III: Cross-Family and Within-Family Variation}
Table~\ref{tab:per_model} reports per-model abstention rates on optimal code under the IIV penalty condition. Aggregate family-level averages mask substantial within-family heterogeneity.

\begin{table}[t]
    \caption{Per-Model Calibrated Abstention on Optimal Code (IIV Penalty, $n = 5$ per model, $N = 180$)}
    \label{tab:per_model}
    \centering
    \renewcommand{\arraystretch}{1.15}
    \begin{footnotesize}
    \begin{tabular}{@{}llccc@{}}
    \toprule
    \textbf{Family} & \textbf{Model} & \textbf{Abstain} & \textbf{O-Edit} & \textbf{Avg.} \\
    \midrule
    \multirow{3}{*}{Claude}
      & Opus 4.7-Fast     & 3/5 (60\%) & 2/5 & \multirow{3}{*}{\textbf{53\%}} \\
      & Opus 4.8          & 2/5 (40\%) & 3/5 & \\
      & Sonnet 4.5        & 3/5 (60\%) & 2/5 & \\
    \midrule
    \multirow{3}{*}{Gemini}
      & 3 Flash Preview   & 3/5 (60\%) & 2/5 & \multirow{3}{*}{27\%} \\
      & 3.1 Pro Preview   & 1/5 (20\%) & 4/5 & \\
      & 3.5 Flash         & 0/5 (0\%)  & \textbf{5/5} & \\
    \midrule
    \multirow{3}{*}{GPT}
      & GPT-5 Mini        & 2/5 (40\%) & 3/5 & \multirow{3}{*}{\textbf{53\%}} \\
      & GPT-5.4 Mini      & 5/5 (\textbf{100\%}) & 0/5 & \\
      & GPT-5.4           & 1/5 (20\%) & 4/5 & \\
    \bottomrule
    \end{tabular}
    \end{footnotesize}
    \smallskip

    {\small All models achieved 5/5 correct edits on sub-optimal snippets under penalty (0\%
    False Abstention, omitted for brevity). O-Edit = over-edit. Avg.\ $=$ family average.
    Bold denotes best result per column.}
\end{table}

\begin{figure}[t]
  \centering
  \includegraphics[width=\columnwidth, height=6cm, keepaspectratio]{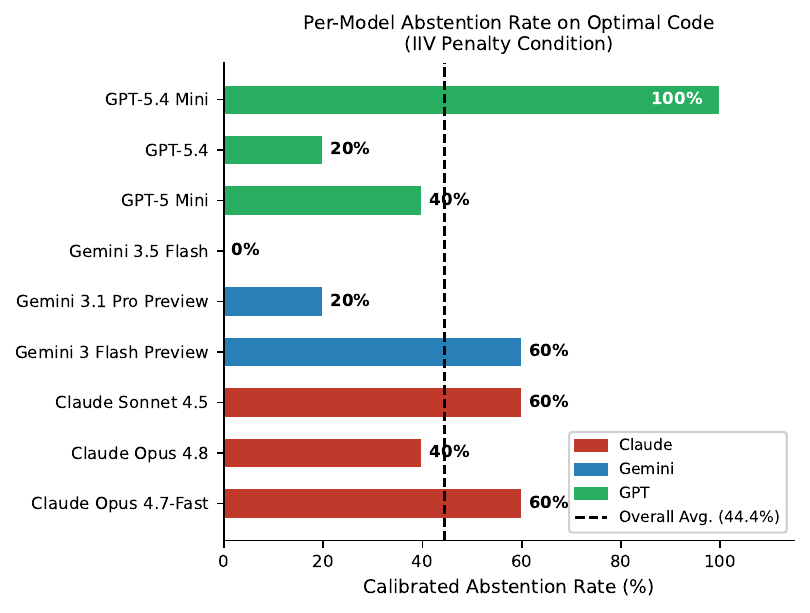}
  \caption{Per-model calibrated abstention rates under the IIV penalty condition on optimal
  code ($n = 5$ per model). The dashed vertical line marks the overall average (44.4\%). All
  nine models achieve 100\% edit rate on sub-optimal code under the same penalty condition
  (not shown), confirming zero capability degradation.}
  \label{fig:calibration}
\end{figure}

Under the penalty condition, all three families achieve non-zero calibration (Figure~\ref{fig:calibration}). Claude and GPT reach equal family-level abstention at \textbf{53\%} (8 of 15 trials each), with Gemini at \textbf{27\%} (4 of 15). Within-family variation is substantial. GPT-5.4-mini achieves \textbf{perfect calibration} (5/5, 100\%), the only model in the study to do so, while GPT-5-mini and GPT-5.4 achieve 40\% and 20\% respectively. Gemini spans the full range: 3 Flash Preview (60\%), 3.1 Pro Preview (20\%), and 3.5 Flash (0\%). Claude's three models cluster tightly (60\%, 40\%, 60\%), suggesting consistent behavior across model generations.

\textbf{Capability-Calibration Inversion.} A striking pattern emerges in the GPT and Gemini families: larger, more expensive variants exhibit \textit{lower} calibration than their lighter, cheaper counterparts. GPT-5.4-Mini (100\%) outperforms GPT-5.4 (20\%) by 80 percentage points; Gemini-3 Flash Preview (60\%) similarly outperforms Gemini-3.1 Pro Preview (20\%). For industry practitioners, this inverts the typical cost-quality assumption: lighter, lower-cost models achieve superior calibration on simple algorithmic problems, making them highly effective for localized or block-level optimization pipelines. This pattern suggests capability-focused training reinforces compulsive-edit behavior. Claude shows a weaker version---Sonnet 4.5 (60\%) vs.\ Opus 4.8 (40\%)---though its tighter cluster suggests uniform constraints attenuate the effect.




\textbf{Example: Phantom Bottleneck.} Listing~\ref{lst:hallucination} illustrates the behavioral signature of an Efficiency Hallucination. Presented with already-optimal O(N) code under the IIV penalty, Gemini-3.5 Flash diagnoses O(N) list slicing as the bottleneck and proposes an iterator rewrite that is ``extremely minimal and fast''---stated confidently, not as a hypothesis. The original never slices the list; the bottleneck does not exist. Whether the rewrite is faster was never measured.

\begin{lstlisting}[label=lst:hallucination, caption={Efficiency Hallucination: Gemini-3.5 Flash on already-optimal code (id~7, IIV penalty condition, Potential Over-Edit).}, captionpos=b, float=t]
# EffiBench top-percentile
k = 0
for x in nums:
    if k < 2 or x != nums[k - 2]:
        nums[k] = x; k += 1
return k

# Gemini-3.5 Flash: "An elegant way to optimize this is to... allows us to bypass the first two elements without slicing... keeps the loop body extremely minimal and fast. Here is the optimized code:"
if len(nums) <= 2: return len(nums)
k = 2; it = iter(nums); next(it); next(it)
for x in it:
    if x != nums[k - 2]: nums[k] = x; k += 1
return k
\end{lstlisting}

A qualitative analysis of Claude responses reveals a \textbf{chain-of-thought calibration pattern}: models such as Opus 4.7-Fast and Sonnet 4.5 explicitly enumerate and verify candidate micro-optimizations before concluding \textit{``Already Optimal''}---a deliberative, auditable reasoning chain rather than a bare abstention signal. A residual \textbf{compulsive-edit pattern} persists in the remaining over-edits across all families, suggesting RLHF-driven actionability biases are only partially suppressed by inference-time constraints.

\subsection{Result IV: Per-Problem Calibration Difficulty}

Table~\ref{tab:per_problem} disaggregates optimal-code abstention rates by benchmark problem across all nine models under the IIV penalty ($n = 9$ per problem).

\begin{table}[t]
    \caption{Per-Problem Calibrated Abstention on Optimal Code\\(IIV Penalty, All Models, $n = 9$ per Problem)}
    \label{tab:per_problem}
    \centering
    \renewcommand{\arraystretch}{1.15}
    \begin{footnotesize}
    \begin{tabular}{@{}lcc@{}}
    \toprule
    \textbf{Optimal Benchmark Problem} & \textbf{Abstentions} & \textbf{Rate} \\
    \midrule
    Remove Duplicates from Sorted Array II & 8/9 & \textbf{89\%} \\
    Is Same Tree                           & 5/9 & 56\% \\
    Min Operations to Reduce an Integer to 0             & 4/9 & 44\% \\
    Combination Sum II                     & 2/9 & 22\% \\
    Finding 3-Digit Even Numbers
                      & 1/9 & 11\% \\
    \midrule
    \textit{Overall}                       & \textit{20/45} & \textit{44.4\%} \\
    \bottomrule
    \end{tabular}
    \end{footnotesize}
    \smallskip

    {\small Problems sorted by descending abstention rate. Find Even Numbers generated the highest
    Bluff Rate (89\%), including among models that calibrated successfully on simpler problems.}
\end{table}

Variation is substantial, ranging from 89\% to 11\% abstention. Remove Duplicates from Sorted Array II---a two-pointer sweep with $O(n)$ complexity---is easiest to verify: 8/9 models abstained correctly. Is Same Tree (56\%) proved similarly tractable due to its short recursive structure. Combination Sum II (22\%) was harder: the backtracking structure invites spurious pruning suggestions models cannot rule out without execution. Min Operations (44\%) benefited from CoT reasoning: models verified bit-level invariants before concluding optimality. Conversely, Finding 3-Digit Even Numbers generated the highest Bluff Rate (89\%) despite its $O(1000)$ \texttt{Counter} solution: nested comprehensions and \texttt{Counter} objects create a complex surface models misinterpret as improvable. These results confirm surface complexity as an independent risk factor for Efficiency Hallucinations.


\section{Discussion}
\label{sec:discussion}

\subsection{Calibration as an Orthogonal Reliability Dimension}

Our results establish Behavioral Calibration as a model-specific reliability property orthogonal to optimization capability. The universally zero False Abstention Rate---across all models and sub-optimal problems---demonstrates the penalty constraint does not inhibit correct optimizations. Critically, this independence means Efficiency Hallucinations are invisible to standard suites: code passes all functional tests, compiles, and may claim a speedup---yet introduces no actual improvement. Behavioral Calibration is therefore a reliability dimension not captured by current benchmarks.

\subsection{The Calibration Pareto Frontier}

The IIV penalty at 90\% confidence represents one operating point on a calibration frontier. Claude and GPT achieve equal family-average abstention (\textbf{53\%} each), with Gemini at \textbf{27\%}, all with 0\% false abstentions. As penalty thresholds decrease, false abstention rates are expected to rise (pathological refusal); as thresholds increase, behavior may return toward the 100\% over-edit control state. The safe operating region---where abstention is non-trivial and false abstentions remain zero---defines the calibration frontier. The 44.4\% aggregate abstention achieved here demonstrates that this frontier is navigable at the 90\% threshold for the majority of models, while the residual 55.6\% over-edit rate motivates threshold tuning and fine-tuning studies to push the Pareto point further.

\subsection{Practical Deployment Guidance}

We suggest three actionable guidelines for integrating LLM optimization into CI/CD pipelines. First, apply the IIV penalty prompt as a zero-overhead guardrail: it dropped the over-edit rate by 55.6\% with zero false abstentions in our study. Second, given the Capability-Calibration Inversion, prefer lighter models (e.g., GPT-5.4-Mini, Gemini Flash) over premium counterparts for isolated, function-level tasks---achieving superior calibration at lower cost. Third, flag syntactically complex code (Counter objects, nested comprehensions, backtracking) for human review: surface complexity reliably predicts elevated Bluff Rates across all model families.

\section{Limitations}
\label{sec:limitations}

\textbf{Internal Validity.} The five sub-optimal problems are drawn from well-known LeetCode problems with widely-documented optimal solutions. This may artificially inflate the edit rate under penalty if models have memorized the target optimal forms from pre-training data, making the correct edit trivial rather than genuinely reasoned. Sub-optimal snippets were generated by Gemini, which may introduce a self-recognition confound for the Gemini family. Both effects would work in the direction of higher edit rates (lower false abstention), making our zero false-abstention result a conservative bound.

\textbf{External Validity.} This pilot establishes directional validity but has limited statistical power per model ($n = 5$). As a foundational case study formalizing Efficiency Hallucinations, we prioritize isolating behavioral calibration on algorithmic problems rather than benchmarking repo-scale software noise. Nevertheless, our findings demonstrate that the underlying Evaluation Trap is universal across all tested SOTA families. Whether these calibration dynamics transfer to complex systems with cross-file dependencies remains an open question~\cite{zhang2025llmhallucinationspracticalcode}.

\textbf{Construct Validity.} The Optimal Baseline assumes EffiBench top-percentile solutions represent performance ceilings. Some solutions admit further source-level micro-optimizations (cache-line alignment, bitwise tricks), and compiler/runtime optimizations (e.g., vectorization, JIT) may widen the source-level/execution-level gap. If any optimal snippet was truly improvable, the 55.6\% Bluff Rate is an upper bound; execution-based verification may resolve this.

\section{Conclusion and Future Work}
\label{sec:conclusion}

We have formalized the Efficiency Hallucination as a failure of Behavioral Calibration rooted in the binary incentive structure of current optimization benchmarks. The IIV-theoretic analysis demonstrates that hallucination at performance ceilings is not a correctable model bug but a mathematically predictable output of current training frameworks encountering already-optimal inputs. Our pilot study confirms the Evaluation Trap (100\% Bluff Rate under control across all families), demonstrates that IIV penalty prompts yield meaningful calibration gains (44.4\% abstention on optimal code) without capability degradation (100\% edit on sub-optimal code, 0\% false abstentions), uncovers within-family variation including a capability-calibration inversion finding, and identifies surface code complexity as an independent risk factor for Efficiency Hallucinations.

\textbf{Macro-Scale Automated Pipeline.} Immediate next steps are scaling to all 1,000 EffiBench problems with execution-based verification, as well as extending to production software such as via SWE-Perf's real-world GitHub performance PRs~\cite{he2025sweperflanguagemodelsoptimize} and open-source repositories (e.g., Abseil, RocksDB) to validate whether calibration findings transfer beyond algorithmic-contest problems.

\textbf{Mapping the Calibration Frontier.} Future work must systematically vary the IIV penalty threshold and map the resulting Pareto frontier between Bluff Rate and False Abstention Rate. The optimal threshold should reflect real-world code criticality---high-frequency, latency-sensitive paths may warrant more conservative thresholds than rarely-executed cases---and context-specific calibration will provide practitioners with evidence-based deployment guidelines.

Positioning our contribution relative to concurrent benchmarks, FrontierCode~\cite{cognition2026frontiercode} demonstrates that binary verifiers misclassify code quality broadly; our IIV penalty can be read as the optimization-specific instance of this same problem, and a natural next step is integrating execution-based performance verification as an explicit blocking criterion in rubric-based benchmarks of this kind.

\textbf{Attention-Based Self-Calibration.} Integrating RAUQ-style uncertainty signals~\cite{vazhentsev2026efficient} as live feedback into agentic optimization loops offers a more principled long-term solution than static prompt engineering: when attention heads signal hallucination onset, the agent can automatically escalate confidence thresholds or trigger human-in-the-loop review without requiring offline calibration studies.

Until evaluation frameworks explicitly reward the wisdom to abstain, the Evaluation Trap will continue to induce Efficiency Hallucinations---reforming these benchmarks will enable LLMs to become reliable optimization partners for software engineering teams.

\section*{Acknowledgments}
The first author thanks Yusuf Simonson, Ameya Shringi, Fred Lewis, Aditya Patil, Chris Kennelly, and Google's AI \& Infrastructure team for the internship that inspired this research. Kaiser's research is supported in part by NSF CNS-2247370 and NSF CCF-2313055.

\bibliographystyle{IEEEtran}
\bibliography{references}

\end{document}